\documentclass{article}
\usepackage{spconf}
\usepackage{IEEEtrantools}
\usepackage[utf8]{inputenc}
\usepackage{cite}
\usepackage{hyperref}
\usepackage{graphicx}
\usepackage{tikz}
	\usetikzlibrary{calc}
	\usetikzlibrary{arrows.meta}
	\usetikzlibrary{shapes}
\usepackage{pgfplots}
	\usepgfplotslibrary{groupplots}
	\pgfplotsset{compat=newest,every axis/.append style={font=\footnotesize}}
\usepackage{amsmath,amssymb,amsthm}
\usepackage{url}
\usepackage{algorithmic}
\usepackage{array}

\title{BONA FIDE RIESZ PROJECTIONS FOR DENSITY ESTIMATION}

\name{Pol {del Aguila Pla}$\dagger\ddagger$ and Michael Unser$\ddagger$}
\address{$\dagger$ CIBM Center for Biomedical Imaging, Switzerland \\ $\ddagger$ Biomedical Imaging Group, École polytechnique fédérale de Lausanne, Lausanne, Switzerland}

    \def\naturals{\mathbb{N}}
    \def\reals{\mathbb{R}}
    \def\integers{\mathbb{Z}}
    
  \def\Ell{\mathrm{L}}
    
    \newcommand{\expect}[1]{\operatorname{\mathrm{E}}\left\lbrace #1 \right\rbrace}
    
    \newcommand{\normal}[2]{\operatorname{\mathcal{N}}\!\left(#1,#2\right)}

    \newcommand{\extrema}[3]{\operatorname{#1}_{#2}\left\lbrace #3 \right\rbrace}
    
    \renewcommand{\min}[2]{\extrema{min}{#1}{#2}}
    
    \def\dint{\mathrm{d}}
    
    \def\analysisfunc{\varphi^{\mathrm{a}}}
    \def\synthesisfunc{\varphi^{\mathrm{s}}}
    
    \def\Vsynthesis{V_{\mathrm{s}}}
    \def\Vanalysis{V_{\mathrm{a}}}
    \def\analysiscoefficients{c_{\mathrm{a}}}
    \def\synthesiscoefficients{c_{\mathrm{s}}}
    \def\correlation{r_{\mathrm{a,s}}}
    \def\projection{\operatorname{P}_{\Vsynthesis}}
    \def\bonafideset{\mathcal{BF}}
    \def\bonafideprojection{\operatorname{P}_{\bonafideset}}
    \def\fdummy{\breve{f}}
    \def\festimation{\tilde{f}}
    
  \theoremstyle{remark}
    
    \newtheorem{lemma}{Lemma}
    
    \newtheorem{definition}{Definition}

\def\thebibliography#1{\section{References}\list
 {[\arabic{enumi}]}{\settowidth\labelwidth{[#1]}\leftmargin\labelwidth
 \advance\leftmargin\labelsep
 \usecounter{enumi}}
 \small
 \def\newblock{\hskip .11em plus .33em minus .07em}
 \sloppy\clubpenalty4000\widowpenalty4000
 \sfcode`\.=1000\relax}

\begin{document}
\bstctlcite{IEEEexample:BSTcontrol}

    
    \maketitle
    
    \begin{abstract}
        The projection of sample measurements onto a reconstruction space represented by a basis on a regular grid is a powerful and simple approach to estimate a probability density function. In this paper, we focus on Riesz bases and propose a projection operator that, in contrast to previous works, guarantees the bona fide properties for the estimate, namely, non-negativity and total probability mass $1$. Our bona fide projection is defined as a convex problem. We propose solution techniques and evaluate them. Results suggest an improved performance, specifically in circumstances prone to rippling effects.
    \end{abstract}
    

    \section{Introduction}
    \label{sec:intro}

The estimation of probability density functions (pdf) pervades most problems
in statistics and machine learning. For instance, the Bayes
classifier achieves optimal classification, but
requires an estimate of the pdf conditioned to each class. 
Similarly, any regression problem can be trivially solved provided
a good estimate of the joint pdf between 
outcomes and covariates is available. 
Practically, pdf estimation
remains one of the most common tools in data
science~\cite{Izenman1991,Takada2008,Wang2019}, with its basic version (a histogram)
being the entry point to any exploratory data analysis.
As a result, the field remains active despite its long 
history~\cite{Uppal2019,Cui2020,Kirkby2021,Ferraccioli2021}.

The mathematical structure of the problem of pdf
estimation is very similar to that of image reconstruction for 
imaging modalities that operate in the limited-photon 
regime, e.g., the construction of a sinogram from positron emission
tomography measurements. From the observation of the empirical
measure $p_\delta$ generated by $N$ independent identically distributed samples $x_n\sim\mathcal{X}$
of a continuous random variable $\mathcal{X}$, with
\begin{IEEEeqnarray}{c} \label{eq:empirical-estimate}
    p_\delta = \frac1N 
    \sum_{n=1}^{N} \delta_{x_n} \,,
\end{IEEEeqnarray}
one aims to recover the probability density function
$f:\reals\rightarrow \reals_+$, of which we assume 
$f\in\Ell_2(\reals)$.

\begin{figure}
	\centering

\begin{tikzpicture}[scale=1.4]
        	\node (f) at (-.2,0) {$p_{\delta}$};
        	\node[draw,rectangle,inner sep=10pt] (prefilter) at (1,0) {$g$};
        	\draw[-latex] (f) -- (prefilter);
        	\node (left_sampl) at (2,0) {\tiny$\bullet$}; \node (right_sampl) at (2.5,0) {\tiny$\bullet$};
        	\draw (prefilter) -- (left_sampl.center) --+ (0.36,0.36);
        	\draw[<->] (2.35,0.1) arc (35:55:1) node[anchor=south] {\tiny $F_\mathrm{s} = \frac1h$};
        	\node[draw, rectangle, dashed, inner sep=10pt] (digital_filter) at (3.5,0) {$q$};
        	\draw[-latex, dashed] (right_sampl.center) -- (digital_filter); 
        	\node (coefficients) at (4.8,0) {$c_\mathrm{s}$};
        	\draw[-latex, dashed] (digital_filter.east) -- (coefficients);
\end{tikzpicture}
	
	\vspace{5pt}
	
	\input{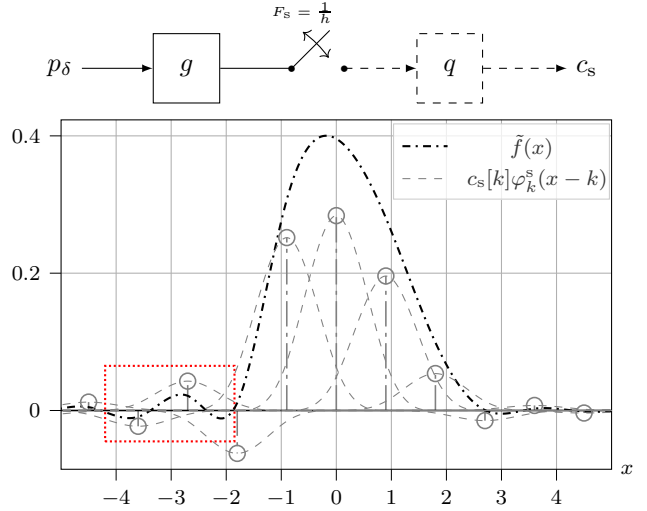}
	
	\vspace{-8pt}
	
    \caption{\small Projection onto the space of uniform splines 
    of degree $3$ of the empirical estimate $p_\delta$ in \eqref{eq:empirical-estimate} for $h=0.9$ and $N=100$
    samples of a standard normal random variable. The projection is implemented by a generalized sampling system composed of a continuous filter 
    $g$ and a digital correction filter $q$. The resulting estimate is 
    not a bona fide pdf, which is the problem we address in this paper.
    \label{fig:sampling}}
\end{figure}

In \cite{Blu2004}, our group proposed a pdf estimator that
relies on the theory of generalized sampling using Riesz bases. However,
in general, the resulting estimates are not bona fide pdfs. 
Although they integrate to
$1$, they are not guaranteed to be nonnegative 
(see Fig.~\ref{fig:sampling}). Recently, 
Cui~\emph{et~al.}~\cite{Cui2020,Kirkby2021} rediscovered the same estimator
and studied it in much detail, but did not provide a technique to 
generate bona fide estimates. In this paper, we present a simple
technique based on convex optimization to obtain better estimates
that are bona fide pdfs within the same framework.

    \section{Sampling and reconstruction} 
    \label{sec:samplingandrec}

A classical problem in signal processing is that of the sampling and reconstruction of continuous-domain signals~\cite{Unser2000,Eldar2015}. 
In short, a function $f\in\Ell_2(\reals)$ is observed through a filter with impulse response $g\in\Ell_2(\reals)$ and sampled
regularly at $x=k h$ for $k\in\integers$. The problem is then to obtain the best approximation 
$\festimation\in\Ell_2(\reals)$ from the collected samples $\analysiscoefficients\in\ell_2$, defined as 
$\analysiscoefficients[k]=(g*f)(kh)$, $\forall k \in \integers$. 
 
An option is to find a projection of $f$ onto a synthesis space
\begin{IEEEeqnarray}{c} \label{eq:Vsynthesis}
 	 \Vsynthesis=   \left\lbrace  \sum_{k\in\integers} \synthesiscoefficients[k] \synthesisfunc_k : 
			 		\synthesiscoefficients \in \ell_2 
				\right\rbrace \subset \Ell_2(\reals)\,,
\end{IEEEeqnarray}
where the synthesis function $\synthesisfunc\in\Ell_2(\reals)$ is scaled and shifted to define 
$\synthesisfunc_k(x) \triangleq\sqrt{1/h} \,\synthesisfunc(x/h-k)$. For convenience, we define the analysis functions $\analysisfunc_k$
\emph{mutatis mutandis} with respect to $\analysisfunc(x)=\sqrt{h}\,g(-hx)$ and note that 
$\analysiscoefficients[k]=\langle \analysisfunc_k, f \rangle$. For the remainder of this paper, 
we assume that 
$\int_{\reals} \analysisfunc(x) \dint x = \int_{\reals} \synthesisfunc(x) \dint x = 1$ and that $\sum_{k\in\integers} \synthesisfunc(x-k) = 1$ (partition of unity).
Provided that $\lbrace \synthesisfunc_k \rbrace_{k\in\integers}$ 
and $\lbrace\analysisfunc_k\rbrace_{k\in\integers}$ are Riesz bases, the coefficients 
$\synthesiscoefficients\in\ell_2$ 
that yield a consistent reconstruction $\festimation\in\Vsynthesis$, in the sense that 
$\langle \analysisfunc_k, \festimation\rangle=\analysiscoefficients[k]$, can be obtained using a discrete
filter~\cite{Unser1994,Unser2000} such that $\synthesiscoefficients[k] = (q*\analysiscoefficients)[k]$, where 
$q$ is the convolutional inverse of the analysis-synthesis correlation sequence 
$\correlation[k] = \langle \analysisfunc_k, \synthesisfunc_0 \rangle$.
This follows directly from the consistent reconstruction condition because
\begin{IEEEeqnarray}{rl} \label{eq:measurement-correlation}
	\langle \analysisfunc_k, \festimation\rangle\, &=
	 \sum_{k'\in\integers} \synthesiscoefficients[k'] \langle \analysisfunc_k, \synthesisfunc_{k'}\rangle =
	 \sum_{k'\in\integers} \synthesiscoefficients[k'] \correlation[k-k'] \nonumber \\
	 &= (\correlation * \synthesiscoefficients)[k]\,. 
\end{IEEEeqnarray}
The function $\festimation$ so constructed is a projection of $f$ onto $\Vsynthesis$, because 
$\projection:\Ell_2(\reals)\rightarrow \Vsynthesis$ 
defined as $f\mapsto\festimation$ fulfills that $\projection\lbrace\festimation\rbrace = \festimation$.
Particularly, if $\Vsynthesis = \Vanalysis$, with $\Vanalysis$ defined analogously to \eqref{eq:Vsynthesis}, 
then $\festimation$ is the minimum $\Ell_2$-norm approximation of 
$f$ in $\Vsynthesis$ (\emph{i.e.}, its orthogonal projection). The polynomial-reproduction properties of $\synthesisfunc$~\cite{Blu1999} then characterize 
the behavior of $\|\projection\lbrace f \rbrace - f\|_{\Ell_2(\reals)}$ as 
$h$ gets small, known as the order of approximation.

    \section{Riesz projections for PDF estimation}
    \label{sec:projection}

\begin{figure}  
    \centering
    \def\xscale{1.4}
\def\yscale{1.4}
\begin{tikzpicture}[xscale=\xscale,yscale=\yscale]
    \draw [->] (0,0) -- (5.2,0) node[anchor=west]{$x$};
    \draw [->] (0,0) -- (0,1.5) node[anchor=south west]{a) $\analysisfunc=\beta^0$};
    \draw[densely dotted] (5,0.2) -- (-0.1,0.2) node[anchor=east]{\tiny$\frac1N$};
    \foreach \x in {1,2,...,4}{
        \draw (\x,-0.05) -- (\x,0);
        \draw (\x-0.5,0) -- (\x-0.5,1) -- (\x+0.5,1) -- (\x+0.5,0);
        \node at (\x,1.2) {\small $\analysisfunc_\x$};
    }
    \foreach \sample in {3.22397672, 2.88117377, 1.74794259, 2.76028579, 2.05813019}{
    	\draw[-latex] (\sample,0) -- (\sample,0.2);
	    \node[red,draw,inner sep=1.5pt,fill=red,fill opacity=0.5] at (\sample,1) {};
    }
    \foreach \x/\y in {1/0,2/0.4,3/0.6,4/0}{
	\draw[blue] (\x,0) -- (\x,\y);
	\node[draw,circle,blue,fill=white, inner sep=1.5pt, fill opacity=0.5] at (\x,\y){};
	\node[blue] at (\x,-0.3) {\small $\analysiscoefficients[\x]$};
    }
\end{tikzpicture} 

\begin{tikzpicture}[xscale=\xscale,yscale=\yscale]
    \draw [->] (0,0) -- (5.2,0) node[anchor=west]{$x$};
    \draw [->] (0,0) -- (0,1.5) node[anchor=south west]{b) $\analysisfunc=\beta^1$};
    \draw[densely dotted] (5,0.2) -- (-0.1,0.2) node[anchor=east]{\tiny$\frac1N$};
    \foreach \x in {1,2,...,4}{
        \draw (\x,-0.05) -- (\x,0);
        \draw (\x-1,0) -- (\x,1) -- (\x+1,0);
        \node at (\x,1.2) {\small $\analysisfunc_\x$};
    }
    \foreach \sample in {3.22397672, 2.88117377, 1.74794259, 2.76028579, 2.05813019}{
	\node[red,draw,inner sep=1.5pt,fill=red,fill opacity=0.5] at (\sample,{\sample-floor(\sample)}){};
	\node[red,draw,inner sep=1.5pt,fill=red,fill opacity=0.5] at (\sample,{-\sample+ceil(\sample)}){};
	\draw[-latex] (\sample,0) -- (\sample,0.2);
    }
    \foreach \x/\y in {1/0.050411482,2/0.409670568,3/0.495122606,4/0.044795344}{
	\draw[blue] (\x,0) -- (\x,\y);
	\node[draw,circle,blue,fill=white, inner sep=1.5pt, fill opacity=0.5] at (\x,\y){};
	\node[blue] at (\x,-0.3) {\small $\analysiscoefficients[\x]$};
    }
\end{tikzpicture} 

\begin{tikzpicture}[xscale=\xscale,yscale=\yscale]
    \draw[-latex] (0,-.05) -- (0,.15);
    \node[anchor=west] at (-.025,0) {\tiny : $\delta_{x_n}$};
    \node[draw,circle,blue,fill=white, inner sep=1.5pt, fill opacity=0.5] (measurement) at (0.7,0) {};
    \node[anchor=west,shift={(-.1,0)}] at (measurement.east) {\tiny : $\analysiscoefficients[k]=\langle p_\delta, \analysisfunc_k\rangle$};
    \node[red,draw,inner sep=1.5pt,fill=red,fill opacity=0.5] (sample) at (2.35,0) {};
    \node[anchor=west,shift={(-.1,0)}] at (sample.east) {\tiny: $ \langle \delta_{x_n}, \analysisfunc_k\rangle = \analysisfunc_k(x_n)$};
    \draw (-0.2,-0.15) rectangle (4.2,0.2);
\end{tikzpicture}
    
    \vspace{-8pt}
    
    \caption{ \small Measurement procedure $\analysiscoefficients[k] = \langle p_\delta, \analysisfunc_k\rangle$ for
    	the cases $\analysisfunc=\beta^0$ and $\analysisfunc=\beta^1$. The coefficient $\analysiscoefficients[k]$ is 
	constructed as an average of the contributions $\analysisfunc_k(x_n)$ of all the samples over the support of $\analysisfunc_k$.
    \label{fig:measurement}}
\end{figure}
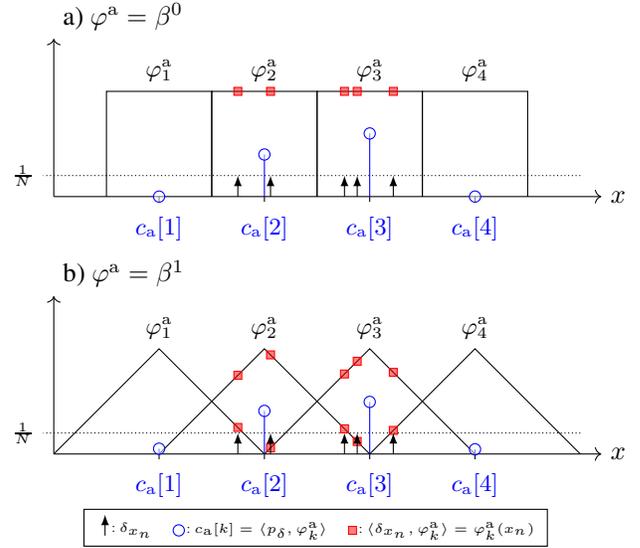

This approach to signal reconstruction can be extended to the estimation of pdfs~\cite{Blu2004}.
In particular, if one chooses $\analysisfunc,\synthesisfunc\in \mathcal{C}_0(\reals)$, where
$\mathcal{C}_0(\reals)$ is the space of continuous functions that decay at infinity 
equipped with the uniform norm $\|\cdot\|_\infty$, then
$\langle \delta_x, \analysisfunc\rangle = \analysisfunc(x)$ is well
defined. Consequently, the approach described above can
be directly applied on the empirical estimate $p_\delta$, as portrayed in Figure~\ref{fig:sampling}, 
to obtain an estimate $\festimation$ of the pdf. For 
$\analysisfunc=\synthesisfunc=\beta^0(x)$, this results in a traditional histogram. 
Here, $\beta^m$ for $m\in\naturals$ is the uniform B-spline of degree $m$.
We illustrate in Figure~\ref{fig:measurement} the measurement process $\langle p_\delta, \analysisfunc_k\rangle$ 
when $\beta^0$ and $\beta^1$ are used as $\analysisfunc$.
In the general case,
Blu and Unser \cite{Blu2004} showed that $\festimation=\projection\left\lbrace p_\delta \right\rbrace$ is an $\Ell_2$-consistent estimator 
for $f$. They also characterized thoroughly its expected $\Ell_2$-error averaged over all possible shifts within $h$
\begin{IEEEeqnarray}{c} \label{eq:error}
	\tilde{\eta}^2 =  \frac1h  \int_{0}^h 
						\expect{\left\|
						 f_\tau - \projection\left\lbrace \frac1N \sum_{n=1}^{N} \delta_{x_n+\tau}\right\rbrace 
						 \right\|^2_{\mathrm{L}_2(\reals)} } \dint\tau. \nonumber \\
\end{IEEEeqnarray}
There, $f_\tau(x) \triangleq f(x-\tau)$ is used for convenience.

Because $\synthesisfunc$ satisfies the partition of unity, we have that $\int_{-\infty}^{\infty} \festimation(x) \dint x=1$. Regretfully, 
this does not guarantee that $\festimation$ will be a bona fide pdf, and we often encounter $\festimation(x)<0$ for some $x\in\reals$.
When $\analysisfunc=\synthesisfunc=\beta^m$ for some $m\in\naturals$, 
the scheme in Figure~\ref{fig:sampling} yields an orthogonal projection. Nonetheless, for degrees $m\geq 1$, the 
 resulting estimate may exhibit negative values due to ripples in the estimate. For example, this is seen inside the dotted box in Figure~\ref{fig:sampling}. 
To address this shortcoming of the method, we propose the following 
projection operator based on an optimization model.

\begin{definition}[Bona Fide Projection]
	The bona fide projection onto $\Vsynthesis$ is
	the operator 
	$\bonafideprojection:\mathcal{M}(\reals) \rightarrow \Vsynthesis$ such that 
	$p_\delta \mapsto \festimation_+$ with
	\begin{IEEEeqnarray}{ll}
		\festimation_+ = & \,\arg \min{\fdummy \in \Vsynthesis}{\| \langle p_\delta, \analysisfunc_k \rangle - 
											\langle \analysisfunc_k, \fdummy\rangle 
										    \|^2_{\ell_2}}\,, \nonumber \\
		&\mbox{ such that } \fdummy(x)\geq 0, \forall x\in\reals\mbox{, and } \int_{-\infty}^{\infty} \fdummy(x) \dint x=1\,. \nonumber  \\ \label{eq:bonafideprojection}
	\end{IEEEeqnarray}
	Here, $\mathcal{M}(\reals)$ is the space of bounded Radon measures. It is the continuous dual of
	$\mathcal{C}_0(\reals)$ and contains Dirac delta distributions.
	Because the set of constraints is convex and the cost function is strictly convex, 
	\eqref{eq:bonafideprojection} has a unique solution.
\end{definition}
We may express succinctly the set of constraints by defining the subset of bona fide pdfs in $\Vsynthesis$ as
$\bonafideset=\lbrace \fdummy \in \Vsynthesis:  \fdummy(x)\geq 0, \forall x\in\reals\mbox{, and } \int_{-\infty}^{\infty} \fdummy(x) \dint x=1\rbrace$.
Lemma~\ref{lem:projection} establishes that \eqref{eq:bonafideprojection} indeed does define a projection
operator, precisely onto $\bonafideset\subset \Vsynthesis$.

\begin{lemma}[Projection] \label{lem:projection}
	The operator $\bonafideprojection$ defined in \eqref{eq:bonafideprojection} is a projection operator onto $\bonafideset$. 
\end{lemma}
\begin{IEEEproof}
	Let $\festimation_+\in\bonafideset$. Because $\bonafideset \subset \Vsynthesis \bigcap \Ell_1(\reals)$ by construction ($\|\festimation_+\|_{\Ell_1}=1$ for any $\festimation_+\in\bonafideset$), we have that  $\bonafideset\subset \mathcal{M}(\reals)$ (see \cite{Unser2020}), and thus
	$\bonafideprojection\lbrace \festimation_+\rbrace$ is well defined. 
	Let $\synthesiscoefficients^{\festimation_+}$ be the unique coefficients of $\festimation_+$ on the basis 
	$\lbrace \synthesisfunc_k \rbrace_{k\in\integers}$, and $\synthesiscoefficients^{\fdummy}$ analogously for $\fdummy$. Then, the cost function in \eqref{eq:bonafideprojection}
	is $\|\correlation * (\synthesiscoefficients^{\festimation_+} - \synthesiscoefficients^{\fdummy})\|_{\ell_2}^2$. Because $\correlation$ has
	a convolutional inverse and $\festimation_+\in\bonafideset$, the unique solution that achieves cost $0$ and fulfills the constraints in \eqref{eq:bonafideprojection} is $\fdummy=\festimation_+$. Therefore, $\bonafideprojection\lbrace \festimation_+\rbrace = \festimation_+$ and
	$\bonafideprojection$ is a projection operator onto $\bonafideset$.
\end{IEEEproof} 

Leveraging \eqref{eq:measurement-correlation}, the optimization problem \eqref{eq:bonafideprojection} can be equivalently stated in terms of the coefficients $\synthesiscoefficients$ of $\fdummy$ as
\begin{IEEEeqnarray*}{l}
\min{\synthesiscoefficients \in \ell_2}{\|\analysiscoefficients - \correlation * \synthesiscoefficients \|^2_{\ell_2} }\\
\mbox{ such that } \sum_{k\in\integers} \synthesiscoefficients[k] \synthesisfunc_k(x) \geq 0, \forall x\in\reals \mbox{ and } 
\sum_{k\in\integers}\synthesiscoefficients[k]=1\,.
\end{IEEEeqnarray*}
However, the enforcement of the non-negativity constraint is
generally a hard problem for all but the simplest bases $\lbrace \synthesisfunc_k \rbrace_{k\in\integers}$. For example, for spline functions
of degree $m$, this entails as many as $m$ semidefinite constraints in the polynomial coefficients describing each 
segment $[k,k+1]$ (see~\cite{Alizadeh2008,Alizadeh2013,Papp2014}). In this paper, we take a general approach that is valid for any 
$\synthesisfunc$ and relies only on linear constraints and convolution. It is inexact but can be made arbitrarily close to 
\eqref{eq:bonafideprojection} at the cost of increased computational complexity. 
Specifically, for a given $M\in\naturals$, 
we impose that $\fdummy(k/M)\geq0$, $\forall k\in\integers$. 
The constrained values are readily computed as 
$\fdummy(k/M) = (\synthesiscoefficients^{\uparrow M} * \varphi^{\mathrm{s},M})[k]$,
where i) $\synthesiscoefficients^{\uparrow M}$ is a sequence of coefficients upsampled from
$\synthesiscoefficients$, so that it contains $(M-1)$ zeros between $\synthesiscoefficients[k]$ and $\synthesiscoefficients[k+1]$,
for every $k\in\integers$, and ii) $\varphi^{\mathrm{s},M}[k] = \synthesisfunc(k/M)$, which corresponds to a short finite-impulse-response filter because $\synthesisfunc$ is often chosen 
with a small support. Therefore, the final optimization problem becomes
\begin{IEEEeqnarray}{l}
\min{\synthesiscoefficients \in \ell_2}{\|\analysiscoefficients - (\correlation * \synthesiscoefficients) \|^2_{2} }\nonumber \\
\mbox{ such that } (\synthesiscoefficients^{\uparrow M} * \varphi^{\mathrm{s},M})[q]\geq 0, \forall q\in\mathbb{Z} \mbox{ and } 
\sum_{k\in\mathbb{Z}}\synthesiscoefficients[k]=1\,. \nonumber \\
\label{eq:final_prob}
\end{IEEEeqnarray}
Problem \eqref{eq:final_prob} has a quadratic-programming structure and can be solved by
a number of standard iterative techniques.

As we shall see in Section~\ref{sec:numres}, and specifically in Figures~\ref{fig:bimodal}~and~\ref{fig:results}, empirical results suggest that 
 the localized nonnegative constraints of \eqref{eq:final_prob} 
lead to estimates that are in $\bonafideset$ and result in smaller approximation errors and a smoother behavior
than the unconstrained solution.
    
    \section{Empirical results} 
    \label{sec:numres}

We implemented $\projection$ as described in Sections~\ref{sec:samplingandrec} 
and~\ref{sec:projection} in \texttt{Python~3.9.5}, leveraging the 
NumPy~\cite{NumPy} and SciPy~\cite{SciPy} libraries for computations.
We implemented an approximated $\bonafideprojection$ as described by
\eqref{eq:final_prob}, which was solved with $M=10$ using \texttt{CVXPY}~\cite{CVXPy}.
An open-source repository that contains all implementations, including those to generate all
the figures in this 
paper, is available through GitHub\footnote{\url{https://github.com/poldap/rpde}}.

\begin{figure}
    \centering
        \input{figs/bimodal}
        
    \vspace{-8pt}
    
    \caption{\small Estimates $\festimation=\projection\lbrace p_\delta \rbrace$ and
    $\festimation_+=\bonafideprojection\lbrace p_\delta \rbrace$ obtained from $N=100$ samples 
    of a Gaussian mixture, $h=0.9$, and
    $\analysisfunc=\synthesisfunc=\beta^3$. Evidently, $\tilde{f}_+$ is a better estimate of $f$. Most importantly, it is also a bona fide pdf. \label{fig:bimodal}}
\end{figure}

A comparison of \cite{Blu2004} and the proposed method on $100$ samples of an equal
mixture of $\normal{3}{1}$ and $\normal{-3}{1}$ with $h=0.9$ and
$\analysisfunc=\synthesisfunc=\beta^3$ results in the estimates
$\tilde{f}=\projection\lbrace p_\delta \rbrace$ and 
$\tilde{f}_+ = \bonafideprojection\lbrace p_\delta \rbrace$ shown in 
Figure~\ref{fig:bimodal}. 
There, one can immediately appreciate that
$\festimation_+$ fulfills the constraints of $\bonafideset$, 
at least within visual tolerance. Furthermore, as highlighted by 
the dotted boxes in the figure, $\festimation_+$ is a much better estimate
overall, exhibiting a much less ripples at the locations
where the true distribution $f$ changes sharply. This 
applies even when the resulting ripples in $\tilde{f}$ do not 
produce negative values (lower-right box). 

\begin{figure}
    \centering
\begin{tikzpicture}

\definecolor{color0}{rgb}{0.12156862745098,0.466666666666667,0.705882352941177}
\definecolor{color2}{rgb}{0.172549019607843,0.627450980392157,0.172549019607843}

\begin{axis}[
width=.495\textwidth,height=.42\textwidth,
legend cell align={left},
legend style={
  fill opacity=0.8,
  draw opacity=1,
  text opacity=1,
  at={(0.6,0.97)},
  anchor=north,
  draw=white!80!black
},
tick align=outside,
tick pos=left,
x grid style={white!69.0196078431373!black},
xlabel={$h$},
x label style={at={(1,0)},anchor=west},
title={Error \emph{vs} Discretization Step},
xmajorgrids,
xmin=0.8,
xmax=1.6,
xminorgrids,
xtick style={color=black},
y grid style={white!69.0196078431373!black},
ylabel={[dB]},
y label style={at={(0,1)},anchor=south, rotate=-90)},
ymajorgrids,
ymin=-25,
ymax=-19.5,
yminorgrids,
ytick style={color=black}
]
\addplot [thick, black, dashdotted]
table {%
0.8 -20.1407817440131
0.837931034482759 -20.4016765663381
0.875862068965517 -20.6539902637598
0.913793103448276 -20.8981703304167
0.951724137931034 -21.1344042049591
0.989655172413793 -21.3625758470338
1.02758620689655 -21.5822139513007
1.06551724137931 -21.79243785239
1.10344827586207 -21.9919091483333
1.14137931034483 -22.1787989875783
1.17931034482759 -22.3507823524708
1.21724137931034 -22.5050713103777
1.2551724137931 -22.6384981323078
1.29310344827586 -22.7476555933888
1.33103448275862 -22.8290948413746
1.36896551724138 -22.8795709277421
1.40689655172414 -22.8963138221528
1.4448275862069 -22.8772915631657
1.48275862068966 -22.8214263238961
1.52068965517241 -22.7287272105084
1.55862068965517 -22.6003165158527
1.59655172413793 -22.4383459977
1.63448275862069 -22.2458205441584
1.67241379310345 -22.0263618947946
1.71034482758621 -21.7839509501951
1.74827586206897 -21.5226835562069
1.78620689655172 -21.2465645953201
1.82413793103448 -20.9593530325347
1.86206896551724 -20.6644597929127
1.9 -20.3648928781382
};
\addlegendentry{Expected $\tilde{\eta}^2$ for $\projection$}
\addplot [semithick, black, mark=*, mark size=3, mark options={solid,fill=black,fill opacity=0.5}, 
]
table {%
0.8 -20.3100225472242
0.837931034482759 -20.239727138274
0.875862068965517 -20.632567726029
0.913793103448276 -20.4218236836711
0.951724137931034 -20.8265102225976
0.989655172413793 -21.3727200983636
1.02758620689655 -21.2357510629462
1.06551724137931 -21.9166406395183
1.10344827586207 -22.0071161101108
1.14137931034483 -22.161119016145
1.17931034482759 -22.5136736542484
1.21724137931034 -22.906431520357
1.2551724137931 -23.5279609610011
1.29310344827586 -23.0085569002851
1.33103448275862 -22.8450679179816
1.36896551724138 -22.5850930183507
1.40689655172414 -22.8370459741316
1.4448275862069 -23.1255846819799
1.48275862068966 -22.8715270678373
1.52068965517241 -22.8855774170221
1.55862068965517 -22.4762710939946
1.59655172413793 -22.5515485574161
1.63448275862069 -22.2002333613623
1.67241379310345 -21.7631558794218
1.71034482758621 -21.9778052965097
1.74827586206897 -21.7457430094458
1.78620689655172 -21.183338160769
1.82413793103448 -20.964467686684
1.86206896551724 -20.7367691076939
1.9 -20.4247175429166
};
\addlegendentry{Empirical $\tilde{\eta}^2$ for $\projection$}
\addplot [semithick, color2, mark=triangle*, mark size=3, mark options={solid,fill=color2,fill opacity=0.5}, 
]
table {%
0.8 -21.2564830526425
0.837931034482759 -20.7690339642818
0.875862068965517 -21.993335696918
0.913793103448276 -21.5098707727954
0.951724137931034 -22.1543225214639
0.989655172413793 -22.5538359771537
1.02758620689655 -23.0250425423226
1.06551724137931 -22.7364400539232
1.10344827586207 -22.973351400977
1.14137931034483 -23.4924029333383
1.17931034482759 -24.5910340602141
1.21724137931034 -24.3799827777844
1.2551724137931 -24.5609192776897
1.29310344827586 -24.3160108685995
1.33103448275862 -24.1621054930128
1.36896551724138 -24.4533323174927
1.40689655172414 -23.8050049331556
1.4448275862069 -23.2826404221747
1.48275862068966 -22.8974432465495
1.52068965517241 -22.1024851020248
1.55862068965517 -21.5713166335106
1.59655172413793 -20.7742934197166
1.63448275862069 -20.2589631514985
1.67241379310345 -19.7345023950528
1.71034482758621 -19.2321984435127
1.74827586206897 -18.6831059800031
1.78620689655172 -18.2489731203309
1.82413793103448 -17.7093355245963
1.86206896551724 -17.305159424343
1.9 -16.8754346456685
};
\addlegendentry{Empirical $\tilde{\eta}^2$ for $\bonafideprojection$}
\end{axis}

\end{tikzpicture}
    
    \vspace{-20pt}
    
    \caption{\small Performance metric $\tilde{\eta}$ in \eqref{eq:error} evaluated for 
    $\projection$ and $\bonafideprojection$ applied to $N=100$ samples of a standard Gaussian.
    Seen here as function of the grid size $h$ and for
    $\analysisfunc=\synthesisfunc=\beta^3$. \label{fig:results}}
\end{figure}
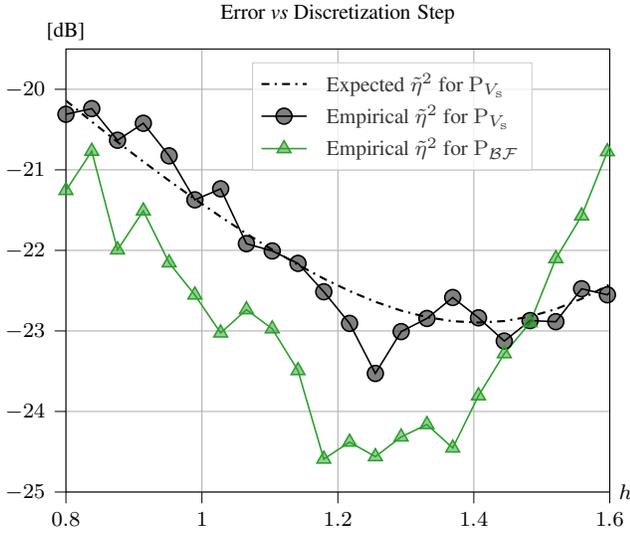

A more detailed study of the expected shift-averaged error incurred by 
both $\festimation$ and $\festimation_+$ is included 
in Figure~\ref{fig:results} and compared to the theoretical predictions
of~\cite{Blu2004} for $\festimation$. 
Both methods are evaluated on $100$ samples of a standard normal
distribution $\normal{0}{1}$ for $h\in[0.8,1.6]$. 
To approximate $\tilde{\eta}$ empirically, the error is computed by
numerical integration over $x$ and is averaged over $120$ realizations. 
For each realization, the data and the probability density function are
shifted by all multiples of $0.025$ between $0$ and $h$, and the results
are averaged to approximate the integral over $\tau$ in \eqref{eq:error}.
The results suggest that $\festimation_+$ improves on 
$\festimation$ by roughly $1~\mathrm{dB}$ for all reasonable values of $h$.
While this improved error is certainly an advantage, the main 
benefit of $\bonafideprojection$ is that its output is directly usable for any application of pdf estimation, because the estimate $\festimation_+\in\bonafideset$ is a bona fide pdf.

    \section{Acknowledgements}
        The authors would like to acknowledge Dr.~Aleix Boquet-Pujadas for
        extensive discussions on the role of non-negativity in spline approximations.
        
        We acknowledge access to the facilities and expertise of the CIBM Center for Biomedical Imaging, a Swiss research center of excellence founded and supported by Lausanne University Hospital (CHUV), University of Lausanne (UNIL), École polytechnique fédérale de Lausanne (EPFL), University of Geneva (UNIGE), and Geneva University Hospitals (HUG).

\bibliographystyle{IEEEtran}
\bibliography{spline_stuff.bib}

\begin{thebibliography}{10}
\providecommand{\url}[1]{#1}
\csname url@samestyle\endcsname
\providecommand{\newblock}{\relax}
\providecommand{\bibinfo}[2]{#2}
\providecommand{\BIBentrySTDinterwordspacing}{\spaceskip=0pt\relax}
\providecommand{\BIBentryALTinterwordstretchfactor}{4}
\providecommand{\BIBentryALTinterwordspacing}{\spaceskip=\fontdimen2\font plus
\BIBentryALTinterwordstretchfactor\fontdimen3\font minus
  \fontdimen4\font\relax}
\providecommand{\BIBforeignlanguage}[2]{{%
\expandafter\ifx\csname l@#1\endcsname\relax
\typeout{** WARNING: IEEEtran.bst: No hyphenation pattern has been}%
\typeout{** loaded for the language `#1'. Using the pattern for}%
\typeout{** the default language instead.}%
\else
\language=\csname l@#1\endcsname
\fi
#2}}
\providecommand{\BIBdecl}{\relax}
\BIBdecl

\bibitem{Izenman1991}
A.~J. Izenman, ``Recent developments in nonparametric density estimation,''
  \emph{Journal of the American Statistical Association}, vol.~86, no. 413, pp.
  205--224, 1991.

\bibitem{Takada2008}
T.~Takada, ``Asymptotic and qualitative performance of non-parametric density
  estimators: {A} comparative study,'' \emph{Econometrics Journal}, vol.~11,
  no.~3, pp. 573--592, 2008.

\bibitem{Wang2019}
Z.~Wang and D.~W. Scott, ``Nonparametric density estimation for
  high-dimensional data—{A}lgorithms and applications,'' \emph{WIREs
  Computational Statistics}, vol.~11, no.~4, p. e1461, 2019.

\bibitem{Uppal2019}
A.~Uppal, S.~Singh, and B.~Poczos, ``Nonparametric density estimation \&
  convergence rates for {GANS} under {B}esov {IPM} losses,'' in \emph{Advances
  in Neural Information Processing Systems}, H.~Wallach \emph{et~al.}, Eds.,
  vol.~32.\hskip 1em plus 0.5em minus 0.4em\relax Curran Associates, Inc.,
  2019.

\bibitem{Cui2020}
Z.~Cui, J.~L. Kirkby, and D.~Nguyen, ``Nonparametric density estimation by
  {B}-spline duality,'' \emph{Econometric Theory}, vol.~36, no.~2, pp.
  250--291, 2020.

\bibitem{Kirkby2021}
J.~L. Kirkby, A.~Leitao, and D.~Nguyen, ``Nonparametric density estimation and
  bandwidth selection with {B}-spline bases: {A} novel {G}alerkin method,''
  \emph{Computational Statistics \& Data Analysis}, vol. 159, p. 107202, 2021.

\bibitem{Ferraccioli2021}
F.~Ferraccioli, E.~Arnone, L.~Finos, J.~O. Ramsay, and L.~M. Sangalli,
  ``Nonparametric density estimation over complicated domains,'' \emph{Journal
  of the Royal Statistical Society: Series B (Statistical Methodology)},
  vol.~83, no.~2, pp. 346--368, 2021.

\bibitem{Blu2004}
T.~Blu and M.~Unser, ``Quantitative {${\mathbf L}^{2}$} approximation error of
  a probability density estimate given by its samples,'' in \emph{Proceedings
  of the Twenty-Ninth {IEEE} International Conference on Acoustics, Speech, and
  Signal Processing ({ICASSP'04})}, vol. {III}, Montr{\'{e}}al QC, CA, May
  17-21 2004, pp. 952--955.

\bibitem{Unser2000}
M.~Unser, ``Sampling---50 {Y}ears after {S}hannon,'' \emph{Proceedings of the
  {IEEE}}, vol.~88, no.~4, pp. 569--587, 2000.

\bibitem{Eldar2015}
Y.~C. Eldar, \emph{Sampling {T}heory, {B}eyond {B}andlimited {S}ystems}.\hskip
  1em plus 0.5em minus 0.4em\relax Cambridge University Press, 2015.

\bibitem{Unser1994}
M.~Unser and A.~Aldroubi, ``A general sampling theory for nonideal acquisition
  devices,'' \emph{{IEEE} Transactions on Signal Processing}, vol.~42, no.~11,
  pp. 2915--2925, 1994.

\bibitem{Blu1999}
T.~Blu and M.~Unser, ``Quantitative {F}ourier analysis of approximation
  techniques: {P}art {I}---{I}nterpolators and projectors,'' \emph{IEEE
  Transactions on Signal Processing}, vol.~47, no.~10, pp. 2783--2795, 1999.

\bibitem{Unser2020}
M.~Unser, ``A note on {BIBO} stability,'' \emph{{IEEE} Transactions on Signal
  Processing}, vol.~68, pp. 5904--5913, 2020.

\bibitem{Alizadeh2008}
F.~Alizadeh, J.~Eckstein, N.~Noyan, and G.~Rudolf, ``Arrival rate approximation
  by nonnegative cubic splines,'' \emph{Operations Research}, vol.~56, no.~1,
  pp. 140--156, 2008.

\bibitem{Alizadeh2013}
F.~Alizadeh and D.~Papp, ``\BIBforeignlanguage{en}{Estimating arrival rate of
  nonhomogeneous {Poisson} processes with semidefinite programming},''
  \emph{\BIBforeignlanguage{en}{Annals of Operations Research}}, vol. 208,
  no.~1, pp. 291--308, 2013.

\bibitem{Papp2014}
D.~Papp and F.~Alizadeh, ``Shape-constrained estimation using nonnegative
  splines,'' \emph{Journal of Computational and Graphical Statistics}, vol.~23,
  no.~1, pp. 211--231, 2014.

\bibitem{NumPy}
C.~R. Harris \emph{et~al.}, ``Array programming with {NumPy},'' \emph{Nature},
  vol. 585, no. 7825, pp. 357--362, 2020.

\bibitem{SciPy}
P.~Virtanen \emph{et~al.}, ``{SciPy} 1.0: {F}undamental algorithms for
  scientific computing in {P}ython,'' \emph{Nature Methods}, vol.~17, pp.
  261--272, 2020.

\bibitem{CVXPy}
S.~Diamond and S.~Boyd, ``{CVXPY}: {A} {P}ython-embedded modeling language for
  convex optimization,'' \emph{Journal of Machine Learning Research}, vol.~17,
  no.~83, pp. 1--5, 2016.

\end{thebibliography}
\end{document}